# Global Existence and Large Time Asymptotic Bounds of $L^\infty$ Solutions of Thermal Diffusive Combustion Systems on $R^n$

P. Collet
Centre de Physique Théorique
Laboratoire CNRS UPR 14
Ecole Polytechnique
F-91128, Palaiseau, France.

and

J. Xin
Department of Mathematics
University of Arizona
Tucson, AZ 85721, USA.

**Abstract**

We consider the initial value problem for the thermal-diffusive combustion systems of the form: $u_{1,t} = \Delta_x u_1 - u_1 u_2^m$, $u_{2,t} = d\Delta_x u_2 + u_1 u_2^m$, $x \in R^n$, $n \geq 1$, $m \geq 1$, $d > 1$, with bounded uniformly continuous nonnegative initial data. For such initial data, solutions can be simple traveling fronts or complicated domain walls. Due to the well-known thermal-diffusive instabilities when $d$, the Lewis number, is sufficiently away from one, front solutions are potentially chaotic. It is known in the literature that solutions are uniformly bounded in time in case $d \leq 1$ by a simple comparison argument. In case $d > 1$, no comparison principle seems to apply. Nevertheless, we prove the existence of global classical solutions and show that the $L^\infty$ norm of $u_2$ can not grow faster than $O(loglogt)$ for any space dimension. Our main tools are local $L^p$ a-priori estimates and time dependent spatially decaying test functions. Our results also hold for the Arrhenius type reactions.

# 1  Introduction

In this paper, we are concerned with the existence of global classical solutions and large time asymptotic bounds of the thermal-diffusive combustion system:

$$\begin{aligned} u_{1,t} &= \Delta_x u_1 - u_1 u_2^m, \\ u_{2,t} &= d\Delta_x u_2 + u_1 u_2^m, \end{aligned} \qquad (1.1)$$

with nonnegative initial data $(u_1, u_2)|_{t=0} = (a_1(x), a_2(x)) \in (C_{b,u}^0(R^n))^2$, the space of uniformly bounded continuous functions on $R^n$. Here $x \in R^n$, $n$, $m$ are positive integers, $d > 1$ is the Lewis number; $\Delta_x$ is the $n$ dimensional Laplacian. System (1.1) describes the evolution of mass fraction of reactant $A$, $u_1$, and that of the product $B$, $u_2$, for the autocatalytic chemical reaction of the form: $A + mB \to (m+1)B$ with rate $k_m u_1 u_2^m$, $k_m$ a positive constant. In case $m = 1$, or 2, we refer to Billingham and Needham, [5], [6], for details. System (1.1) also describes the mass fraction, $u_1$, and temperature, $u_2$, of reactant $A$, of a one step irreversible reaction $A \to B$; especially when $u_2^m$ is replaced by the Arrhenius reaction term $\exp\{-E/u_2\}$, $E > 0$ being the activation energy. In this context, system (1.1) is the well-known thermal diffusive system, see Matkowsky and Sivashinsky [19].

One of the basic questions for (1.1) with $L^\infty$ initial data is the existence of global solutions and the possible uniform in time bounds of $u_2$. In case of the Arrhenius reaction, i.e. with $u_1 \exp\{-\frac{E}{u_2}\}$ replacing $u_1 u_2^m$ in (1.1), there are many works on global solutions, see Avrin [2], Larrouturou [15] for results in one space dimension, among others. Yet their bounds of the solutions grow linearly in time. It is still a conjecture whether $u_2$ is bounded uniformly in time, see Berestycki and Larrouturou [3], and Manley, Marion, and Temam [16].

On the other hand, system (1.1) on a bounded domain $\Omega$ in $R^n$ with homogeneous boundary conditions has been thoroughly studied. The problem on existence and uniform bounds of solutions was first posed by R. Martin, and later solved partially by Alikakos [1], and completely by Masuda [18]. See also Haraux and Youkana [13], Hollis, Martin, and Pierre [14] for related approaches and extensions. System (1.1) on the line with spatially decaying data in $L^1 \cap L^\infty(R^1)$ has been investigated recently in Berlyand and Xin [4], Bricmont, Kupiainen, and Xin [9] for critical nonlinearity $m = 2$. Global classical solutions exist for any size initial data and converge to self-similar solutions with anomalous exponents [9].

System (1.1) with $L^\infty$ data on $R^n$ is very different from either the one on the bounded domains or the one with spatially decaying data. The system admits propagating front solutions, from simple traveling wave solutions to the complicated domain walls. When $m = 1, 2$, existence of traveling fronts are proved in [5]; in [6], formal asymptotic as well as computational studies are done for fronts generated from initial data $u_1 =$ positive constant, $u_2 =$ bounded nonnegative function with compact support. In the Arrhenius reaction and high activation energy limit ($E \to +\infty$), it is well-known that planar fronts are subject to (thermal-diffusive) instabilities when $d$ is far enough away from one, and fronts become chaotic, see Clavin [10], Sivashinsky [20], Terman [21] and references therein. In the long wave asymptotic limit, the perturbations of the planar fronts satisfy the celebrated Kuramoto-Sivashinsky equation [20]. For an interesting study on stable and unstable planar fronts away from the large $E$ limit, see Bonnet et al. [7]. In spite of the front instabilities, one still has uniformly bounded solutions if $d < 1$. This fact is easy to demonstrate by a simple comparison argument, Martin and Pierre [17]. However, when $d > 1$, no comparison argument seems to apply, and a completely different approach is necessary.

Our method is to seek local $L^p$ estimates in space by studying certain localized nonlinear functionals of solutions. Similar functionals appeared first in [18], and later in [13]. Since our solutions are only bounded in maximum norm and have no spatial decay at infinities, the functionals in [18] and [13] are not directly applicable. As in Collet and Eckmann [12], and Collet [11], we introduce smooth cut-off functions and convert the global functionals of previous authors into local ones. The first kind of cut-off functions we employ are simply: $\varphi = \varphi(x) = (1 + |x - x_0|^2)^{-n}$, where $x_0$ is an arbitrary point in $R^n$, and is used to translate the location of cut-off so as to achieve uniform $L^\infty$ bounds in space. With such a $\varphi$ and the resulting local $L^p$ estimates, we prove the existence of global classical solutions. However, the $L^\infty$ norm of solutions grow exponentially in time. To improve the $L^\infty$ estimates of solutions, we consider the second kind of cut-off functions which are time dependent solutions of the backward heat equation $\varphi_t + d\Delta\varphi = 0$. Using these time dependent cut-off functions, we are able to refine the $L^\infty$ estimates to the order of loglog growth for any space dimension. Thus the possible growth of $u_2$ is practically extremely hard to observe even if it exists. On the other hand, it remains an interesting problem to prove or disprove the uniform $L^\infty$ bounds on $u_2$. Our main result is:

**Theorem 1.1** *Consider the thermal diffusive combustion system (1.1) with nonnegative initial data $(u_1, u_2)|_{t=0} = (a_1, a_2) \in (C^0_{b,u}(R^n))^2$, and $d > 1$. Then there exist unique global in time classical solutions $(u_1, u_2) \in C([0, +\infty); (C^0_{b,u}(R^n))^2) \cap C^1((0, +\infty); (C^0_{b,u}(R^n))^2)$.*

Moreover, let $||(a_1, a_2)||_\infty \equiv \max(||a_1||_\infty, ||a_2||_\infty)$, then there exists positive constant $C = C(n, d, m)$ such that:

$$\begin{aligned} ||u_1(t,x)||_{L^\infty(R^n)} &\leq ||a_1||_\infty, \\ ||u_2(t,x)||_{L^\infty(R^n)} &\leq C||(a_1, a_2)||_\infty loglog(||(a_1, a_2)||_\infty^m t + 2e). \end{aligned} \quad (1.2)$$

**Corollary 1.1** *Under the same assumptions in the above theorem except that the nonlinear term $u_1 u_2^m$ is replaced by either $u_1 exp\{-\frac{E}{u_2}\}$ or $u_1 u_2^m exp\{-\frac{E}{u_2}\}$, $E > 0$, then there exists positive constant $C = C(n, d, ||(a_1, a_2)||_\infty, m)$ such that:*

$$\begin{aligned} ||u_1(t,x)||_{L^\infty(R^n)} &\leq ||a_1||_\infty, \\ ||u_2(t,x)||_{L^\infty(R^n)} &\leq Cloglog(t + 2e). \end{aligned} \quad (1.3)$$

**Remark 1.1** *In case of power nonlinearities $u_1 u_2^m$, the system has the following scale invariance property: if $u_i = u_i(t, x)$, $i = 1, 2$, are solutions, then so are $v_i = v_i(t, x) \equiv \lambda^{\frac{2}{m}} u_i(\lambda^2 t, \lambda x)$, for any $\lambda > 0$. That is why the estimates are also in the scale invariace form in the theorem. In case of the Arrhenius reactions, we lose the scale invariance due to the exponential term $exp\{-\frac{E}{u_2}\}$, hence we do not know the explicit dependence of $C$ on $||(a_1, a_2)||_\infty$. The proofs are the same in both cases except for some technical details that we will point out later.*

**Remark 1.2** *If the initial data for $u_2$, i.e. the function $a_2$ is strictly above any positive constant, then maximum principle shows that $u_2$ is above this constant forever, and so $u_1$ decays to zero exponentially fast. By Theorem 1.1, $u_2$ is uniformly bounded for all time since $u_1 u_2^m$ decays exponentially in time.*

The rest of the paper is organized as follows. In section 2, we use the first kind of time independent cut-off functions and local nonlinear functionals to prove the existence of global solutions. In section 3, we employ the time dependent cut-off functions and their properties to refine the $L^\infty$ estimates of solutions and complete the proof of the main theorem.

## 2 Global Existence of Classical Solutions

In this section, we establish the global existence of the classical solutions of the thermal diffusive system:

$$u_{1,t} = \Delta_x u_1 - u_1 u_2^m,$$

$$u_{2,t} = d\Delta_x u_2 + u_1 u_2^m, \tag{2.1}$$

where $x \in R^n$, $t > 0$, $n \geq 1$, $m \geq 1$, $d > 1$; the initial data $(a_1(x), a_2(x))$ are bounded uniformly continuous functions on $R^n$, denoted by $C_{b,u}^0(R^n)$. Local existence of nonnegative classical solutions on a maximal existence interval $[0, T_0)$ is standard, and we only need to derive estimates of solutions independent of $T_0$, so as to continue the classical solutions forever in time.

Consider classical solutions $u_i \in C([0, T_0); C_{b,u}^0(R^n)) \cap C^1((0, T_0); C_{b,u}^0)$, $i = 1, 2$, for some $T_0 > 0$. Let $F = F(u_1, u_2)$ be a smooth function of $u_i$, such that $F \geq 0$, $F_i \geq 0$, $F_{i,i} \geq 0$, $i = 1, 2$, here we abbreviate $F_i = \frac{\partial F}{\partial u_i}$, similarly for the second derivatives. Let also $\varphi = \varphi(t, x)$ be a smooth nonnegative function with exponential spatial decay at infinity. Writing $\int$ in place of $\int_{R^n} \cdots dx$, we calculate using (2.1) and integration by parts:

$$\begin{aligned}
\partial_t \int \varphi F &= \int \varphi_t F + \int \varphi F_1 u_{1,t} + \int \varphi F_2 u_{2,t} \\
&= \int \varphi_t F + \int \varphi F_1 \Delta u_1 + d \int \varphi F_2 \Delta u_2 - \int \varphi (F_1 - F_2) u_1 u_2^m \\
&= \int \varphi_t F - \int \varphi F_{1,1} |\nabla u_1|^2 - (d+1) \int \varphi F_{1,2} \nabla u_1 \cdot \nabla u_2 - d \int \varphi F_{2,2} |\nabla u_2|^2 \\
&\quad - \int F_1 \nabla \varphi \cdot \nabla u_1 - d \int F_2 \nabla \varphi \cdot \nabla u_2 - \int \varphi (F_1 - F_2) u_1 u_2^m. \tag{2.2}
\end{aligned}$$

In view of:

$$- \int F_1 \nabla \varphi \cdot \nabla u_1 - \int F_2 \nabla \varphi \cdot \nabla u_2 = \int \Delta \varphi F,$$

we get:

$$\begin{aligned}
\partial_t \int \varphi F &= \int (\varphi_t + d\Delta\varphi) F + (d-1) \int F_1 \nabla \varphi \cdot \nabla u_1 \\
&\quad - \int \varphi [F_{1,1} |\nabla u_1|^2 + (1+d) F_{1,2} \nabla u_1 \cdot \nabla u_2 + d F_{2,2} |\nabla u_2|^2] \\
&\quad - \int \varphi (F_1 - F_2) u_1 u_2^m, \tag{2.3}
\end{aligned}$$

which is our basic identity.

By maximum principle, $\|u_1\|_\infty(t) \leq \|a_1\|_\infty \leq C < +\infty$; $u_i \geq 0$, with strict inequality for $t > 0$, any $x \in R^n$. To apply (2.3), we require:

$$F_1 \geq 2F_2,$$
$$(d+1)^2 F_{1,2}^2 \leq d F_{1,1} F_{2,2}, \tag{2.4}$$

for any $(u_1, u_2) \in [0, C] \times R^+$. Under the conditions (2.4), we have from (2.3):

$$\partial_t \int \varphi F \leq \int (\varphi_t + d\Delta\varphi) F + (d-1) \int F_1 \nabla\varphi \cdot \nabla u_1$$
$$- \frac{1}{2} \int \varphi [F_{1,1}|\nabla u_1|^2 + dF_{2,2}|\nabla u_2|^2]$$
$$- \frac{1}{2} \int \varphi F_1 u_1 u_2^m. \tag{2.5}$$

As a first application of (2.5), we choose:

$$\varphi = \varphi(x) = \frac{1}{(1+|x-x_0|^2)^n},$$
$$F(u_1, u_2) = (A + u_1 + u_1^2) e^{\epsilon u_2}, \tag{2.6}$$

where $x_0$ is an arbitrary point in $R^n$ so that we can translate the function $\varphi$ to achieve uniform estimates of solutions in space; and $A$, $\epsilon^{-1}$ suitably large to be determined. We verify conditions (2.4) as follows:

$$F_1 = (1 + 2u_1) e^{\epsilon u_2},$$

$$F_2 = \epsilon(A + u_1 + u_1^2) e^{\epsilon u_2},$$

so:

$$F_1 \geq 2F_2, \quad \text{for} \quad (u_1, u_2) \in [0, C] \times R^1, \quad \text{if} \quad 2\epsilon(A + C + C^2) < 1. \tag{2.7}$$

Also:

$$F_{1,1} = 2e^{\epsilon u_2},$$

$$F_{2,2} = \epsilon^2 (A + u_1 + u_1^2) e^{\epsilon u_2},$$

$$F_{1,2} = \epsilon(1 + 2u_1) e^{\epsilon u_2},$$

thus:

$$\frac{(d+1)^2 F_{1,2}^2}{dF_{1,1}F_{2,2}} = \frac{(1+2u_1)^2(d+1)^2}{2d(A+u_1+u_1^2)} \leq \frac{(1+2C)^2(d+1)^2}{2dA} < 1,$$
$$\text{if} \quad A > \frac{(1+2C)^2(d+1)^2}{2d}. \tag{2.8}$$

Combining (2.7) and (2.8), we see that for any given $C$ and $d$, we first choose $A$ according to (2.8) then $\epsilon$ by (2.7). It follows from (2.5) for $t \in [0, T_0)$:

$$\partial_t \int \varphi F \leq \int d\Delta\varphi F + (d-1) \int F_1 \nabla\varphi \cdot \nabla u_1 - \frac{1}{2} \int \varphi F_{1,1} |\nabla u_1|^2. \tag{2.9}$$

Now $\varphi$ has the properties:
$$|\Delta\varphi| \leq K\varphi, \quad |\nabla\varphi| \leq K\varphi, \tag{2.10}$$

for some constant $K > 0$. We continue from (2.9):

$$\begin{aligned}
\partial_t \int \varphi F &\leq \int dK\varphi F + (d-1)K \int F_1\varphi|\nabla u_1| - \frac{1}{2}\int \varphi F_{1,1}|\nabla u_1|^2 \\
&\leq dK\int \varphi F + \frac{1}{2}(d-1)^2 K^2 \int \varphi \frac{F_1^2}{F_{1,1}}.
\end{aligned} \tag{2.11}$$

Notice that:
$$\frac{F_1^2}{F_{1,1}} = \frac{(1+2u_1)^2 e^{\epsilon u_2}}{2} \leq 2(A + u_1 + u_1^2)e^{\epsilon u_2} = 2F, \text{ since } A > 1. \tag{2.12}$$

Finally we end up with:
$$\partial_t \int \varphi F \leq [dK + (d-1)^2 K^2] \int \varphi F,$$

or:
$$\int \varphi F \leq \Gamma e^{\sigma t}, \tag{2.13}$$

where
$$\sigma = dK + (d-1)^2 K^2, \quad \Gamma \leq c(n)(A + C + C^2)e^{\|a_2\|_\infty}, \tag{2.14}$$

where $c(n)$ is a positive dimensional constant.

We now prove that inequality (2.13) implies that for any unit cube $Q$ and any finite $p \geq 1$:
$$\int_Q |u_2|^p \leq A^{-1}\Gamma e^{\sigma t} 2^n \epsilon^{-p}(p+1)^{p+1}. \tag{2.15}$$

In fact, we have with any nonegative integer $k$:
$$e^{\sigma t}\Gamma \geq \int \varphi F \geq A \int \varphi e^{\epsilon u_2} \geq \epsilon^k A \int_Q \varphi \frac{u_2^k}{k!} \geq \frac{A\epsilon^k 2^{-n}}{k!} \int_Q u_2^k, \tag{2.16}$$

by taking $x_0$ at the center of $Q$. Hence,
$$\int_Q |u_2|^k \leq A^{-1}\Gamma e^{\sigma t} 2^n \epsilon^{-k} k!, \tag{2.17}$$

which implies (2.15) by interpolation.

Now we look for $L^\infty$ bound on $u_2$. Let
$$G_\tau(z) = (4\pi d\tau)^{-\frac{n}{2}} e^{-\frac{|z|^2}{4d\tau}}, \quad z \in R^n,$$

then
$$u_2(t,x) = G_t \star a_2 + \int_0^t G_{t-s} \star (u_1 u_2^m)(s) ds. \tag{2.18}$$

The first term on the right hand side of (2.18) is bounded by $||a_2||_\infty$. Moreover,

$$G_{t-s} \star (u_1 u_2^m)(s,x) = (4\pi d(t-s))^{-\frac{n}{2}} \int e^{-\frac{(x-y)^2}{4d(t-s)}} (u_1 u_2^m)(s,y) dy.$$

Let $\{Q_j\}$, $j = 0, 1, 2, \cdots$, be the tiling of $R^n$ by unit cubes $Q_j$'s such that $x$ is at the center of $Q_0$. We have:

$$\int e^{-\frac{(x-y)^2}{4d(t-s)}} (u_1 u_2^m)(s,y) dy = \sum_{Q_j} \int_{Q_j} e^{-\frac{(x-y)^2}{4d(t-s)}} (u_1 u_2^m)(s,y) dy. \tag{2.19}$$

For $y \in Q_j$, we have the inequality:

$$e^{-\frac{(x-y)^2}{8d(t-s)}} \leq \sup_{y \in Q_j} e^{-\frac{(x-y)^2}{8d(t-s)}} = e^{-\frac{\text{dist}(x,Q_j)^2}{8d(t-s)}}.$$

Also there exists positive dimensional constant $c = c(n)$ such that if $y \in Q_j$, $j \neq 0$, we have:

$$c(n) \text{dist}^2(x, Q_j) \geq |x-y|^2. \tag{2.20}$$

Applying Hölder's inequality with $p > \frac{n}{2}$ and its conjugate $q$, we get:

$$\int_{Q_j} e^{-\frac{(x-y)^2}{8d(t-s)}} (u_1 u_2^m)(s,y) dy \leq (\int_{Q_j} e^{-\frac{q(x-y)^2}{8d(t-s)}} dy)^{1/q} \cdot (\int_{Q_j} (u_1^p u_2^{mp})(s,y) dy)^{1/p}$$
$$\leq c(n) d^{n/2q} (t-s)^{n/2q} (\int_{Q_j} (u_1^p u_2^{mp})(s,y) dy)^{1/p}$$
$$\leq (t-s)^{n/2q} \Omega(n,d,a_i,t), \tag{2.21}$$

where $c(n)$ is a dimensional constant and by (2.15):

$$\Omega(n,d,a_i,t) = c(n) d^{n/2q} ||a_1||_\infty [A^{-1} \Gamma e^{\sigma t} 2^n \epsilon^{-mp} (mp+1)^{mp+1}]^{1/p}$$
$$= c(n,d,a_i) e^{\sigma t/p} \epsilon^{-m} (mp+1)^{m+1/p}, \tag{2.22}$$

here $c(n,d,a_i)$ is a positive constant depending on $n$, $d$, and $||(a_1,a_2)||_\infty$.

We deduce from (2.18)——(2.22) that:

$$G_{t-s} \star u_1 u_2^m(s,x) \leq (4\pi d(t-s))^{-n/2} \sum_{Q_j} e^{-\frac{\text{dist}(x,Q_j)^2}{8d(t-s)}} \int_{Q_j} e^{-\frac{(x-y)^2}{8d(t-s)}} (u_1 u_2^m)(s,y) dy$$

$$\leq (4\pi d)^{-n/2}\Omega(n,d,a_i,t)(t-s)^{-n/2p}\sum_{Q_j} e^{-\frac{dist(x,Q_j)^2}{8d(t-s)}}$$

$$\leq (4\pi d)^{-n/2}\Omega(n,d,a_i,t)(t-s)^{-n/2p}(\int_{R^n} e^{-\frac{|x-y|^2}{8dc(n)(t-s)}}dy+1)$$

$$\leq (4\pi d)^{-n/2}\Omega(n,d,a_i,t)(t-s)^{-n/2p}(c(n,d)(t-s)^{n/2}+1)$$

$$\leq c(n,d)\Omega(n,d,a_i,t)((t-s)^{n/2q}+(t-s)^{-n/2p})). \quad (2.23)$$

So integrating (2.23) on $s \in [0,t]$ gives:

$$||u_2(t,x)||_\infty \leq ||a_2||_\infty + c(n,d)\Omega(n,d,a_i,t)(t^{\frac{n}{2q}+1}+t^{1-\frac{n}{2p}}), \quad (2.24)$$

where $p > n/2$. Estimate (2.24) and the standard parabolic regularity theory then implies the global classical solution $(u_1,u_2)(t,x) \in (C([0,+\infty);C^0_{u,b}) \cap C^1((0,+\infty);C^0_{u,b}))^2$.

## 3  Large Time Asymptotic Bounds of Solutions

In this section, we improve the $L^\infty$ estimates (2.24) from exponentially growing in time to the order of loglog growth and complete the proof of the main theorem. Let us choose as before:

$$F = F(u_1,u_2) = (A+u_1+u_1^2)e^{\epsilon u_2},$$

yet the function $\varphi$ is now a solution of the backward heat equation:

$$\varphi_t + d\Delta\varphi = 0.$$

Define:

$$g(u) = A + u + u^2,$$

and

$$\tilde{g} = u + u^2.$$

Let us consider $t \in [0,T)$, where $T$ is a suitably large but fixed time. The function $\varphi$ is explicit:

$$\varphi = \varphi(t,T;x) = (4\pi d)^{-\frac{n}{2}}(T-t)^{-\frac{n}{2}}e^{-\frac{|x|^2}{4d(T-t)}}. \quad (3.1)$$

With the above choice, inequality (2.5) gives:

$$\partial_t \int \varphi F \leq (d-1)\int F_1(\nabla\varphi\cdot\nabla u_1) - \frac{1}{2}\int \varphi[F_{1,1}|\nabla u_1|^2 + dF_{2,2}|\nabla u_2|^2] - \frac{1}{2}\int \varphi F_1 u_1 u_2^m. \quad (3.2)$$

The first integral of the right hand side of (3.2) can be transformed using integration by parts as follows:

$$\int F_1(\nabla\varphi \cdot \nabla u_1) = \int g'(u_1)e^{\epsilon u_2}(\nabla\varphi \cdot \nabla u_1) = \int e^{\epsilon u_2}\nabla\varphi \cdot \nabla g(u_1)$$
$$= \int e^{\epsilon u_2}\nabla\varphi \cdot \nabla \tilde{g}(u_1) = -\int \Delta\varphi e^{\epsilon u_2}\tilde{g}(u_1) - \epsilon\int e^{\epsilon u_2}\tilde{g}(u_1)\nabla\varphi \cdot \nabla u_2$$
$$= J_1 + J_2. \tag{3.3}$$

In view of (3.1), we see that:

$$\Delta\varphi = -d^{-1}\varphi_t = -d^{-1}(4\pi d)^{-\frac{n}{2}}[\frac{n}{2}(T-t)^{-(\frac{n}{2}+1)} - \frac{|x|^2}{4d(T-t)^{2+\frac{n}{2}}}]e^{-\frac{|x|^2}{4d(T-t)}}$$
$$\geq -\frac{n}{2d(4\pi d)^{n/2}}\frac{1}{(T-t)^{\frac{n}{2}+1}}e^{-\frac{|x|^2}{4d(T-t)}}$$
$$\geq -\frac{c(n,d)}{T-t}\varphi, \tag{3.4}$$

which implies that:

$$J_1 \leq \frac{c(n,d)}{T-t}\int \varphi\tilde{g}(u_1)e^{\epsilon u_2}. \tag{3.5}$$

On the other hand,

$$|J_2| \leq \int \frac{\tilde{g}^2(u_1)|\nabla\varphi|^2}{g(u_1)\varphi}e^{\epsilon u_2} + \frac{\epsilon^2}{4}\int \varphi g(u_1)|\nabla u_2|^2 e^{\epsilon u_2}. \tag{3.6}$$

Combining (3.2)—(3.6), we get ($F_{2,2} = \epsilon^2 g(u_1)e^{\epsilon u_2}$):

$$\partial_t\int \varphi F \leq (d-1)\frac{c(n,d)}{T-t}\int \varphi\tilde{g}(u_1)e^{\epsilon u_2} + (d-1)\int \frac{\tilde{g}^2(u_1)|\nabla\varphi|^2}{g(u_1)\varphi}e^{\epsilon u_2}$$
$$- \frac{1}{2}\int \varphi g'(u_1)u_1 u_2^m e^{\epsilon u_2}$$
$$= c(n,d)\int \varphi e^{\epsilon u_2}[\frac{\tilde{g}(u_1)}{T-t} + \frac{\tilde{g}^2(u_1)|\nabla\varphi|^2}{g(u_1)\varphi^2} - g'(u_1)u_1 u_2^m]. \tag{3.7}$$

Notice that:

$$\nabla\varphi = \frac{1}{(4\pi d)^{n/2}} \cdot \frac{1}{(T-t)^{\frac{n}{2}}} \cdot \frac{-x}{2d(T-t)} \cdot e^{-\frac{|x|^2}{4d(T-t)}}$$
$$= -c(n,d)\frac{x}{T-t}\varphi,$$
$$\frac{|\nabla\varphi|^2}{\varphi^2} = \frac{c(n,d)|x|^2}{(T-t)^2}, \tag{3.8}$$

which along with (3.7) yields:

$$\begin{aligned}\partial_t \int \varphi F &\leq c(n,d) \int \varphi e^{\epsilon u_2}[\frac{u_1+u_1^2}{T-t} + \frac{(u_1+u_1^2)^2|x|^2}{(A+u_1+u_1^2)(T-t)^2} - (1+2u_1)u_1 u_2^m] \\ &\leq c(n,d,a_i) \int \varphi e^{\epsilon u_2} u_1(\frac{1}{T-t} + \frac{|x|^2}{(T-t)^2} - u_2^m) \\ &\leq c(n,d,a_i) \int_{\{x\in R^n | \frac{u_2^m}{2} \leq \frac{1}{T-t} + \frac{|x|^2}{(T-t)^2}\}} \varphi e^{\epsilon u_2} u_1(\frac{1}{T-t} + \frac{|x|^2}{(T-t)^2} - \frac{u_2^m}{2}) \\ &\quad - \frac{c(n,d,a_i)}{2} \int \varphi e^{\epsilon u_2} u_1 u_2^m \\ &\leq c(d,n,a_i) \int_{R^n} \varphi u_1 e^{2\frac{1}{m}\epsilon(\frac{1}{T-t}+\frac{|x|^2}{(T-t)^2})^{\frac{1}{m}}} (\frac{1}{T-t} + \frac{|x|^2}{(T-t)^2}) dx \\ &\quad - \frac{c(n,d,a_i)}{2} \int \varphi e^{\epsilon u_2} u_1 u_2^m. \end{aligned} \qquad (3.9)$$

**Remark 3.1** *If we have the Arrhenius type reactions $g_{ar} \equiv u_2^m exp\{-\frac{E}{u_2}\}$, which is monotonely increasing in $u_2$, we simply need to replace the explicit inverse $(g_{ar})^{\frac{1}{m}}$ by an implicit one $g_{ar}^{-1}$. Everything else is the same.*

If $T - t \geq 1$, then requiring $\epsilon < \frac{1}{8d}$ when $m = 1$, we get:

$$\begin{aligned}\partial_t \int \varphi F &\leq \frac{c(n,d,a_i)}{T-t} \int (T-t)^{-\frac{n}{2}} e^{-\frac{|x|^2}{4d(T-t)}} (1 + \frac{|x|^2}{T-t}) e^{2\epsilon(\frac{1}{T-t}+\frac{|x|^2}{(T-t)^2})^{\frac{1}{m}}} \\ &\quad - \frac{c(n,d,a_i)}{2} \int \varphi e^{\epsilon u_2} u_1 u_2^m \\ &\leq \frac{c(n,d,a_i)}{T-t} \int (T-t)^{-\frac{n}{2}} e^{-\frac{|x|^2}{4d(T-t)}} (1 + \frac{|x|^2}{T-t}) e^{2\epsilon(1+\frac{|x|^2}{T-t})^{\frac{1}{m}}} \\ &\quad - \frac{c(n,d,a_i)}{2} \int \varphi e^{\epsilon u_2} u_1 u_2^m \\ &\leq \frac{c(n,d,a_i)}{T-t} - \frac{c(n,d,a_i)}{2} \int \varphi e^{\epsilon u_2} u_1 u_2^m, \end{aligned}$$

which implies that:

$$\begin{aligned}\frac{c(n,d,a_i)}{2} \int_0^t \int \varphi e^{\epsilon u_2} u_1 u_2^m + \int \varphi F &\leq \int \varphi(0,T;x) F(a_1,a_2) dx + c(n,d) log T \\ &= ||F(a_1,a_2)||_\infty + c(n,d) log T, \end{aligned} \qquad (3.10)$$

for $t \in [0, T-1)$. It follows from (3.10) that:

$$\int_0^{T-1} dt \int_{R^n} \varphi e^{\epsilon u_2} u_1 u_2^m dx \leq c(n,d,a_i)(1 + log T) \qquad (3.11)$$

By choosing $\varphi = \varphi(t, T; x - x_0)$, for any $x_0 \in R^n$, we also arrive at (3.11) and so:

$$\int_0^{T-1} dt \varphi \star (e^{\epsilon u_2} u_1 u_2^m)(t) \leq c(n, d, a_i)(1 + logT). \tag{3.12}$$

Let $v = u_2^k$, $k \geq 2$, then

$$v_t = k u_2^{k-1} u_{2,t},$$

$$\nabla v = k u_2^{k-1} \nabla u_2,$$

$$\Delta v = k u_2^{k-1} \Delta u_2 + k(k-1) u_2^{k-2} |\nabla u_2|^2,$$

and so $v$ satisfies the equation:

$$v_t = d\Delta v - dk(k-1) u_2^{k-2} |\nabla u_2|^2 + k u_1 u_2^{m+k-1}, \tag{3.13}$$

which implies:

$$v_t \leq G_t \star v_0 + k \int_0^t ds G_{t-s} \star (u_1 u_2^{k+m-1})(s, x). \tag{3.14}$$

Letting $t = T$ in (3.14) yields:

$$v(T, x) \leq ||a_2||_\infty^k + k \int_0^{T-1} ds G_{T-s} \star (u_1 u_2^{k+m-1})(s) + k \int_{T-1}^T ds G_{T-s} \star (u_1 u_2^{k+m-1})(s),$$

which shows by (3.12):

$$v(T, x) \leq ||a_2||_\infty^k + c(n, d, a_i) \epsilon^{-k} k!(1 + logT) + k \int_{T-1}^T ds G_{T-s} \star (u_1 u_2^{k+m-1})(s). \tag{3.15}$$

Now it suffices to consider:

$$\int \varphi(t, T; x) u_1 u_2^{k+m-1}(x),$$

where $t \in [T-1, T]$.

Let $\{Q_j\}_0^\infty$ be a tiling of $R^n$ with unit cubes, and 0 located at the center of $Q_0$. Then:

$$\int \varphi(t, T; x) u_1 u_2^{k+m-1}(x) \leq c(n, d) \sum_{Q_j} \int_{Q_j} \frac{1}{(T-t)^{\frac{n}{2}}} e^{-\frac{|x|^2}{4d(T-t)}} u_1 u_2^{k+m-1} dx$$

$$\leq c(n, d) \sum_{Q_j} \frac{1}{(T-t)^{\frac{n}{2}}} e^{-\frac{dist(0, Q_j)^2}{8d(T-t)}} \int_{Q_j} e^{-\frac{|x|^2}{8d(T-t)}} u_1 u_2^{k+m-1} dx. \tag{3.16}$$

Following the argument in (2.19) and (2.21), we get:

$$\int_{Q_j} e^{-\frac{|x|^2}{8d(T-t)}} u_1 u_2^{k+m-1} dx \leq c(n, d)(T-t)^{\frac{n}{2q}} (\int_{Q_j} u_2^{p(k+m-1)}(t, y) dy)^{\frac{1}{p}}. \tag{3.17}$$

On the other hand, adjusting $T$ to $T+1$ in (3.10), we have for $t \in [T-1, T]$:

$$\int \varphi(t, T+1; x) F(u_1, u_2) dx \leq c(n, d, a_i)(1 + log(T+1)),$$

or

$$\int e^{-\frac{|x|^2}{4d(T+1-t)}} e^{\epsilon u_2} dx \leq c(n, d, a_i)(1 + logT),$$

or

$$\int e^{-\frac{|x|^2}{4d}} e^{\epsilon u_2} dx \leq c(n, d, a_i)(1 + logT).$$

Using the spatially translated $\varphi$, we find:

$$\int_{Q_j} e^{\epsilon u_2} dx \leq c(n, d, a_i)(1 + logT), \quad \forall \; j,$$

or

$$\int_{Q_j} u_2^\gamma dx \leq \gamma! c(n, d, a_i) \epsilon^{-\gamma}(1 + logT), \quad \forall \; \gamma, \; j, \; \in Z^+, \; t \in [T-1, T). \tag{3.18}$$

By (3.18), (3.17) gives by Hölder's inequality:

$$\int_{Q_j} e^{-\frac{|x|^2}{8d(T-t)}} u_1 u_2^{k+m-1} dx \leq c(n, d, a_i)(T-t)^{\frac{n}{2q}} \left(\int_{Q_j} u_2^{[\beta]+1} dy\right)^{\frac{\beta}{p([\beta]+1)}},$$

$$\leq c(n, d, a_i)(T-t)^{\frac{n}{2q}} (([\beta]+1)!)^{\frac{\beta}{p([\beta]+1)}} \epsilon^{-\frac{\beta}{p}} (1 + logT)^{\frac{\beta}{p([\beta]+1)}}, \tag{3.19}$$

where $\beta = p(k+m-1)$, $[\beta]$ stands for the integral part of $\beta$.

**Proof of Theorem 1.1:** Combining (3.16) and (3.19) and again following the argument in (2.19)——(2.24) shows for $T \geq e$:

$$\int \varphi(s, T; x) u_1 u_2^{k+m-1}(s, x) dx$$

$$\leq c(n, d, a_i)((T-s)^{\frac{n}{2q}} + (T-s)^{-\frac{n}{2p}})(([\beta]+1)!)^{\frac{\beta}{p([\beta]+1)}} \epsilon^{-\frac{\beta}{p}} (logT)^{\frac{\beta}{p([\beta]+1)}}, \tag{3.20}$$

where $p > \max(1, \frac{n}{2})$. Integrating (3.20) from $T-1$ to $T$ shows via (3.15) that:

$$v(T, x) \leq ||a_2||_\infty^k + c(n, d, a_i)\epsilon^{-k}(logT)k! + c(n, d, a_i)\epsilon^{-\frac{\beta}{p}}((p(k+m))!)^{1/p}(logT)^{1/p},$$

or:

$$u_2(T, x) \leq ||a_2||_\infty + c(n, d, a_i)(k! logT + (logT)^{1/p}((p(k+m))!)^{1/p})^{1/k},$$

by Sterling's formula ( $(p(k+m))! \leq c(n, m) e^{pk log k}$):

$$\leq c(n, d, a_i, m)(e^{k log k} e^{log log T} + e^{log log T} e^{k log k})^{1/k}$$

$$\leq c(n,d,a_i,m)e^{\frac{loglogT}{k}+logk}. \tag{3.21}$$

Minimizing the exponent with respect to $k$ shows that we should choose:

$$k = [loglogT] + 1, \tag{3.22}$$

which implies from (3.21), and (2.24) that:

$$u_2(T,x) \leq c(n,d,a_i,m)loglog(T+2e) \tag{3.23}$$

for all $T \geq 0$, where constant $c > 0$ depends on $n$, $d$, $||(a_1,a_2)||_\infty$, and $m$. In case of power nonlinearities $u_1 u_2^m$, we first consider initial data such that $||(a_1,a_2)||_\infty \leq 1$ and so drop the dependence on $a_i$ in the bound (3.23). We verify by direct substitution that if $u_i(t,x)$, $i=1,2$, are solutions, then for any $\lambda > 0$, $\lambda^{\frac{2}{m}} u_i(\lambda^2 t, \lambda x)$ are also solutions. Now choose $\lambda$ such that

$$\lambda^{-\frac{2}{m}} = ||(a_1,a_2)||_\infty. \tag{3.24}$$

It follows that the $L^\infty$ norm of the initial data of the solutions $\lambda^{\frac{2}{m}} u_i(\lambda^2 t, \lambda x)$ is equal to one. By (3.23), we have:

$$\lambda^{\frac{2}{m}}||(u_1,u_2)(\lambda^2 t, \lambda x)||_\infty \leq c(n,d,m)loglog(t+2e), \tag{3.25}$$

which implies:

$$||(u_1,u_2)(\lambda^2 t, x)||_\infty \leq c(n,d,m)\lambda^{-\frac{2}{m}}loglog(t+2e).$$

Rescaling time $t$, we obtain:

$$||(u_1,u_2)(t,x)||_\infty \leq c(n,d,m)\lambda^{-\frac{2}{m}}loglog(\lambda^{-2}t+2e),$$

which is just:

$$||(u_1,u_2)(t,x)||_\infty \leq c(n,d,m)||(a_1,a_2)||_\infty loglog(||(a_1,a_2)||_\infty^m t + 2e), \tag{3.26}$$

by recalling (3.24). The proof of the main theorem is complete.

## Acknowledgements.

The work was done when both authors were visiting the Institut Mittag-Leffler (IML) during the Fall of 1994. We thank IML for its hospitality and pleasant environment, and A. Kupiainen for organizing the workshop on Nonlinear PDE's, Turbulence, and Statistical Mechanics. P. C. thanks M. Benedicks and the Mathematics Department of the Kungliga

Tekniska Högskolan for their kind hospitality and financial support. The work of J. X. was partially supported by the Swedish Natural Science Research Council (NFR) grant F-GF 10448-301, and the US National Science Foundation grant DMS-9302830.